\documentclass[conference]{IEEEtran}
\IEEEoverridecommandlockouts
\usepackage{amsmath,epsfig}
\usepackage{bm}
\usepackage{slashbox}
\usepackage{algorithmic}
\usepackage{latexsym}
\usepackage{amsmath}
\usepackage{amssymb}
\usepackage{epsfig}
\usepackage{multicol}
\usepackage{graphicx}
\usepackage[ruled,vlined]{algorithm2e}
\usepackage{epstopdf}
\usepackage{subfigure}
\usepackage{wrapfig}
\usepackage{floatflt}
\usepackage{graphicx,subfigure}
\usepackage{amsmath, amsthm, amssymb}
\usepackage[ruled,vlined]{algorithm2e}
\usepackage{epstopdf}

\newcommand{\HH}{\mathbf{H}}

\newcommand{\z}{\mathbf{z}}

\newcommand{\M}{\mathbf{M}}
\newcommand{\I}{\mathbf{I}}

\newcommand{\rr}{\mathbf{r}}

\begin{document}

\title{{Bad Data Injection Attack and Defense in Electricity Market using Game Theory Study}}

\author{\authorblockN{Mohammad Esmalifalak$^\dag$, Ge Shi$^\ddag$, Zhu Han$^\dag$, and Lingyang Song$^\ddag$ \\
\authorblockA{$^\dag$ECE Department, University of Houston, Houston, TX 77004\\
$^\ddag$School of Electrical Engineering and Computer Science, Peking University, Beijing, China}}
}

\maketitle\thispagestyle{empty}

\begin{abstract}

Applications of cyber technologies improve the quality of monitoring and decision making in smart grid. These cyber technologies are vulnerable to malicious attacks, and compromising them can have serious technical and economical problems. This paper specifies the effect of compromising each measurement on the price of electricity, so that the attacker is able to change the prices in the desired direction (increasing or decreasing). Attacking and defending all measurements are impossible for the attacker and defender, respectively. This situation is modeled as a zero--sum game between the attacker and defender. The game defines the proportion of times that the attacker and defender like to attack and defend different measurements, respectively.  From the simulation results based on the PJM 5-Bus test system, we can show the effectiveness and properties of the studied game.

\end{abstract}

\section{Introduction}
\begin{figure}
\centering
\includegraphics[width=90mm]{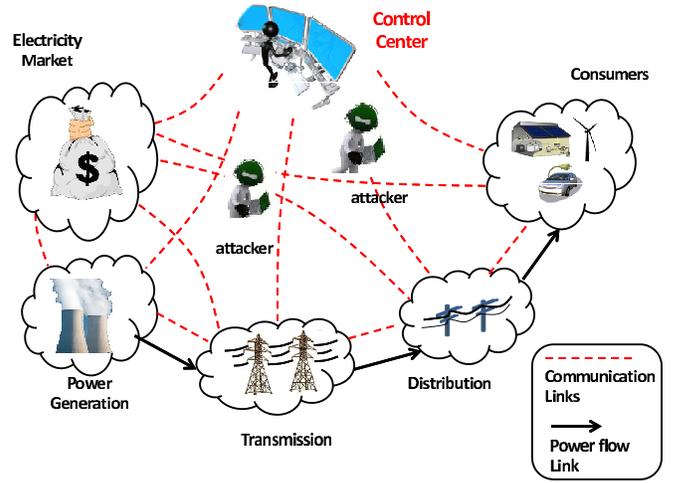}
\caption{Flow of energy and data between different parts of smart grids }
\label{f:smartgrid}
\end{figure}

Recently, power systems are becoming more and more sophisticated in the structure and configuration because of the increasing in electricity demand and the limited energy resources. Traditional power grids are commonly used to carry power from a few central generators to a large number of customers. In contrast, the new-generation of electricity grid that is also known as the smart grid uses bidirectional flows of electricity and information to deliver power in more efficient ways responding to wide ranging conditions and events \cite{TFGA-2008} (Fig. \ref{f:smartgrid}).

Online monitoring of smart grid is important for control centers in different decision making processes. State estimation (SE) is a key function in building real-time models of electricity networks in Energy Management Centers (EMCs) \cite{Monti}. State estimators provide precise and efficient observations of operational constraints to identify the current operating state of the system in quantities such as transmission line loadings or bus voltage magnitudes. Accuracy of state estimation can be affected by bad data during the measuring process.
Measurements may contain errors due to the various reasons such as random errors, incorrect topology information and injection of bad data by attackers. By integrating more advanced cyber technologies into the energy management system (EMS), cyber-attacks can cause major technical problems such as blackouts in power systems\footnote{Aurora attack involves a cyber attack against breakers in a generating unit. This experiment  shows the abilities of cyber attackers in taking control over breakers and consequently, it reveals the technical problems of this attack for the power grid \cite{Meserve}.}\cite{KEN1,KEN2}. The attacks also can be designed to the attacker's financial benefit at the expense of the general consumer's net cost of electricity \cite{WCNCesmali, xie_li}.

In this paper, we consider the case wherein the attacker uses cyber attack against electricity prices. We show that the attacker observes the results of the day--ahead market and changes the estimated transmitted power in order to change the congestion\footnote{Injected power in a specific node of power network, will be transferred to different loads through transmission lines (using kirchoff's law). In power community we say congestion happens if increasing the power injection, increases (at least one of) transmission lines' power to their (its) thermal limit \cite{Kumar_Srivastava,MALAesmali}.} level, resulting in a profit. On the other hand, the defender tries to defend the accuracy of network measurements. Since the attacker and defender are not able to attack and defend all measurements, they will compete to increase and decrease the injected false data, respectively. This behavior is modeled by a two-person zero-sum strategic game where the players try to find the Nash equilibrium and maximize their profits. The results of simulations on the PJM 5-Bus test system show the effectiveness of attack on the prices of electricity on the real--time market.

The remainder of this paper is organized as follows: The literature survey is provided in Section \ref{LIT:SURV}. The system model is given
in Section \ref{sec:model}, and the formulation of an undetectable attack in the electricity market is given in Section \ref{sec:attinmarket}. Section \ref{sec:game} models the interactions between the attacker and defender as a zero--sum game. Numerical results are shown in Section \ref{res:list}, and
the conclusion closes the paper in Section \ref{sec:conclusion}.

\section{Literature Review}
\label{LIT:SURV}
Due to the importance of the smart grid studies, some surveys have classified the different aspects of smart grids \cite{XFSM-,E-Brown,Rohjansand}. In \cite{XFSM-} the authors explore three major systems, namely the smart infrastructure system, the smart management
system, and the smart protection system and also propose possible future directions in each system. In \cite{E-Brown}, a survey is designed to define a "smart distribution system" as well as to study the implications of the smart grid initiative on distribution engineering. In \cite{Rohjansand} relevant approaches are investigated to give concrete recommendations for smart grid standards, which try to identify standardization in the context of smart grids. National Institute of Standards and Technology (NIST) in \cite{NIST_Roadmap}, explains anticipated benefits and requirements of smart grid.

Some researches have been done over cyber security for smart grid \cite{Liu, Esmali_ICA, AMRE-2010-1,Husheng_Li, GCZD-2011,NIST_DRAFT}. In \cite{Liu}, an undetectable attack by bad data detectors (BDD) is first introduced, where the attacker knows the state estimation Jacobian matrix ($H$) and defines an undetectable attack using this matrix.
\cite{Esmali_ICA} uses independent component analysis (ICA), and inserts an undetectable attack even when this matrix is unknown for attackers.
In \cite{AMRE-2010-1}, the authors discuss key security technologies for a smart grid system, including public key infrastructures and trusted computing. Reliable and secure state estimation in smart grid from communication capacity requirement point of view is analyzed in \cite{Husheng_Li}.
In \cite{GCZD-2011}, a new criterion of reliable strategies for defending power systems is derived and two allocation algorithms have been developed to seek reliable strategies for two types of defense tasks. \cite{NIST_DRAFT} is a draft from NIST which addresses the cyber security of smart grid extensively. While most of current researches (in bad data injection area) focus on different attack or defend scenarios, our work describes a mutual interaction between both parties. This work shows how the interest of one party (attacker or defender) can influence the other's interest.

Some applications of game theory in smart grids have been studied in \cite{Swearingen,ZhuLambotharan,Fadlullah,BuYuLiu}.
In \cite{Swearingen}, the authors present a method for evaluating a fully automated electric grid in real time and
finding potential problem areas or weak points within the electric grid by using the game theory. In \cite{ZhuLambotharan}, the authors propose a consumption scheduling mechanism for home and neighborhood area load demand management in smart grid using integer linear programming (ILP) and game theory. \cite{Fadlullah} is a survey about some of game theory-based applications to solve different problems in
smart grid. In \cite{BuYuLiu} the authors model and analyze the interactions between the retailer and electricity customers
as a four-stage Stackelberg game.

Demand-side management (DSM), is another topic in smart grid, which is recently considered by researchers.
In \cite{ZhangLu} an intelligent management system is designed based on the objective of orderly consumption and
demand-side management, under the circumstances of China's smart grid construction.
An Intelligent Metering/Trading/Billing System (ITMBS) with its
implementation on DSM is analyzed by \cite{WangHuang}.  \cite{MohsenianWong} is a research on an autonomous and distributed demand-side energy management system among different users.

\section{System Model}
\label{sec:model}
In power systems, transmission lines are used to transfer
generated power from generating units to consumers. Theoretically, transmitted complex power between
bus $i$ and bus $j$ depends on the voltage difference between these two buses, and it is also a function of impedance between these buses. In general, transmission lines have high reactance over resistance (i.e. $X/R$ ratio), and one can
approximate the impedance of a transmission line with its reactance. In DC power flow studies, it is assumed that the voltage phase difference
between two buses is small and that the amplitudes of voltages in buses are near to
 unity. Transmitted power is approximated with a linear equation ~\cite{woll}:
\begin{equation}\label{active.power.flow}
P_{ij}=\frac{\theta_i-\theta_j}{X_{ij}},
\end{equation}
where $\theta_i$ is the voltage phase angle in bus $i$, and $X_{ij}$
is the reactance of transmission line between bus $i$ and bus $j$.
In the state-estimation problem, the control center tries  to estimate $n$ phase angles $\theta_i$, by observing $m$ real-time measurements. In power flow studies, the voltage phase angle ($\theta_i$) of the reference bus is fixed and known,
and thus only $n-1$ angles need to be estimated.  We define the state vector as $\mathbf{\theta}=[\theta_1,\dots, \theta_n]^T.$
The control center observes a vector $\mathbf{z}$ for $m$ active power measurements. These measurements can be either transmitted active power $P_{ij}$ from bus $i$ to $j$, or injected active power to bus $i$ ($P_{i}=\sum P_{ij}$).  The observation can be
described as follows:
\begin{equation}\label{actual.power.flow.matrix}
\mathbf{z}=\mathbf{P}( \mathbf{\theta})+\mathbf{e},
\end{equation}
where $\mathbf{z}=[z_{1},\cdots,z_{m}]^T$ is the vector of measured active power in transmission lines,
$\mathbf{P}(\mathbf{\theta})$ is the nonlinear relation between measurement $z$, state $\mathbf{\theta}$ is the vector of $n$ bus phase angles $\theta_i$, and $\mathbf{e}=[e_1,\cdots,e_m]^T$ is the
Gaussian measurement noise vector with covariant matrix $\mathbf{\Sigma}_e$.

Define the Jacobian matrix $\mathbf{H} \in \mathbb{R}^{m}$ as
\begin{equation}
\mathbf{H}=\frac{\partial \mathbf{P}({\mathbf{\theta}})}{\partial \mathbf{\theta}}\mid_{\mathbf{\theta}=\,\mathbf{0}}.
\end{equation}
If the phase difference ($\theta_i-\theta_j$) in (\ref{active.power.flow}) is small, then the linear approximation model of (\ref{actual.power.flow.matrix}) can be described as:
\begin{equation}
\label{linear.power.flow.matrix}
\mathbf{z}= \mathbf{H} \mathbf{\theta} +  \mathbf{e}.
\end{equation}
The bad data can be injected to $\textbf z$ so as to influence the state estimation of $\theta$.
Next, we describe the current bad data injection method used in state estimators of different electricity markets. Given the power flow measurements $\mathbf{z}$, the estimated state vector $\hat{\mathbf{\theta}}$ can be computed as:
\begin{equation}
\hat{\mathbf{\theta}}=(\mathbf{H}^T \mathbf{\Sigma}_e^{-1}\mathbf{H})^{-1}\mathbf{H}^T \mathbf{\Sigma}_e^{-1}\mathbf{z}=\mathbf{M}\mathbf{z},
\label{eq:state_infer}
\end{equation}
where
\begin{equation}\label{EQ:M}
\M=(\mathbf{H}^T \mathbf{\Sigma}_e^{-1}\mathbf{H})^{-1}\mathbf{H}^T \mathbf{\Sigma}_e^{-1}.
\end{equation}
 Thus, the residue vector $\mathbf r$ can be computed as the difference between measured quantity and the calculated value from the estimated state:
\begin{equation}\label{EQ:resid}
\rr = \z - \HH\hat{\mathbf{\theta}}.
\end{equation}
Therefore, the expected value and the covariance of the
residual are:
\begin{equation}
E(\rr) =  0 \ \mbox{and} \
cov(\rr)  =  (\I-\M)\mathbf{\Sigma}_e,\label{eq:variance}
\end{equation}

False data detection can be
performed using a threshold test ~\cite{AAAG-2004}. The hypothesis of not being attacked is accepted if
\begin{equation}
\max_i{|r_i|} \le \gamma,
\label{eq:rn_test}
\end{equation}
where $\gamma$ is the threshold and $r_i$ is the component of $\mathbf r$.

\section{Attack in electricity market}\label{sec:attinmarket}
A power network is a typically large and complicated system, which should be operated without any interruption. Normal operation  needs a system wide monitoring of the states of network in specific time intervals. Based on the monitored values, corrective actions need to be taken. Any fault in measurement data (because of measurement failures or cyber attack against them), can change the decisions of control center, which can cause serious technical or economical problems in the network. In this section, we first introduce the electricity market structure, and then from the attacker point of view we will formulate an undetectable attack that can change the prices of electricity.
\subsection{Optimal Power Flow (OPF) and DCOPF}\label{elecmarket}
Security and optimality of power network operation are the most important tasks in control centers, which can be achieved by efficient monitoring and decision making. After deregulation of electric industries, different services that can improve security and optimality of network can be traded in different markets. Energy market is one of these markets in which generation companies (GENCO's) and load serving entities (LSE's) compete to generate and consume energy, respectively\footnote{In an electricity (energy) market, GENCO's submit their bids (for generating electricity) to the market. In this case, higher prices will decrease the chance of supplying electricity (selling electricity). Similarly, LSE's submit their bids for consuming energy. In this case, lower bids will decrease the chance of buying electricity. So competition in both entities (GENCO's and LSE's) will increase the efficiency of the electricity market.}. Control center knowing the submitted prices and network constraints, tries to maximize social welfare for all participants. A well known program for solving this optimization is Optimal Power Flow (OPF) program. Linear form of optimal power flow is called (DCOPF) and is used to define the price of electricity (called locational marginal prices or LMPs) in both day--ahead and real--time markets. In the following subsections, the formulation of DCOPF together with the general structure of day--ahead and real--time markets is described.

\subsection{DC Optimal Power Flow (DCOPF)}

In general, the LMP can be split
 into three components including the marginal energy price $LMP_i^{Energy}$, marginal congestion price $LMP_i^{Cong}$, and marginal loss price $LMP_i^{Loss}$ \cite{FLRB-2007,FLJP-2004,ELTZ-2004}. A common model of the LMP simulation is introduced in \cite{FLRB-2007}. It is based on the DC model and Linear Programming (LP), which can easily incorporate both marginal congestion and marginal losses. The generic dispatch model can be written as
\begin{eqnarray}\label{DCOPFNL_EA}
\min_{\mathbf{G_i}} \quad  \quad \sum\limits_{i=1}^{N} C_i \times G_i,
\end{eqnarray}
\[\mbox{s.t.}\ \left\{
\begin{array}{l l}
\sum\limits_{i=1}^{N} G_i - \sum\limits_{i=1}^{N} D_i = 0,\\
\sum\limits_{i=1}^{N} GSF_{k-i} \times (G_i - D_i) \leq F^{max}_k,  \; k \in \{all \; lines\},\\
G^{min}_i \leq G_i \leq G^{max}_i, \,  \; i \in \{all \; generators\},
\end{array}
\right.
\]
where
\begin{alignat*}{2}
& N & & \text{number of buses;}\\
& C_i & & \text{generation cost at bus $i$ in (\$/MWh);}\\
& G_i & & \text{generation dispatch at bus $i$ in (\$/MWh);}\\
& D_i & & \text{demand at bus $i$ in (MWh);}\\
& GSF_{k-i} &\ & \text{generation shift factor from bus $i$ to line $k$;}\\
& F_k^{max} & & \text{transmission limit of line $K$;}\\
& G_i^{max} & & \text{upper generation limit for generator $i$;}\\
& G_i^{min} & & \text{lower generation limit for generator $i$.}\\
\end{alignat*}

The general formulation of the LMP at bus $i$ can be written as follows:
\begin{alignat}{1}
& LMP_i = LMP^{energy} + LMP^{cong}_i +LMP^{loss}_i,\\
& LMP^{energy} = \lambda,\\
& LMP^{cong}_i = \sum\limits_{i=1}^{L} GSF_{k-i} \times \mu_k,\\
& LMP^{loss}_i = \lambda \times (DF_i - 1),\label{FORDCOPL}
\end{alignat}
where $L$ is the number of lines, $\lambda$ is the Lagrangian multiplier of the equality constraint, $\mu_k$ is the Lagrangian multiplier of the $k^{th}$ transmission constraint, and $DF_i$ is delivery factor at bus $i$. If the optimization model in (\ref{DCOPFNL_EA}) ignores losses, we will have $DF_i = 1$ and $LMP_i^{loss} = 0$ in (\ref{FORDCOPL}). In this work in order to emphasize the main point to be presented, the loss price is ignored.

\subsubsection{Day-Ahead Market}
 Based on the submitted bids (from generators and loads) and predicted network condition\footnote{Such as the load level for the next day, which can be predicted by the historical load data from the past years.}, control center runs the DCOPF program. The output of this market specifies the dispatch schedule for all generators and defines the Locational Marginal Price (LMP) in each bus of power network. Trading electricity in most of electricity markets such as PJM Interconnection, New York, and New England markets is based on the LMP method.

\subsubsection{Real--Time Market} In this market the control center  conducts the following: 1- Gathers data from the measurements that are installed in the physical layer (power network); 2- Estimates the states of the network (online monitoring of the network); 3- Runs an incremental dispatch model based on the state estimation results. The obtained LMP's will be considered as the real-time price of electricity\footnote{Dispatch schedule will be similar to the day--ahead market and major changes of load will be covered by the Ancillary Services.}.  The real--time (Ex--Post) model which is used in Midwest ISO, PJM, and ISO-New England, can be written as \cite{LOTT,TZHLI}:
\begin{eqnarray}\label{DCOPFNL_EP}
\min_{\mathbf{\Delta G_i}} \quad  \quad \sum\limits_{i=1}^{N} C_i^{RT} \times \Delta G_i,
\end{eqnarray}
\[\mbox{s.t.}\ \left\{
\begin{array}{l l}
\sum\limits_{i=1}^{N} \Delta G_i - \sum\limits_{i=1}^{N} \Delta D_i = 0,\\
\sum\limits_{i=1}^{N} GSF_{k-i} \times (\Delta G_i - \Delta D_i) \leq 0,  \; k \in \{CL\},\\
\Delta G^{min}_i \leq \Delta G_i \leq \Delta G^{max}_i, \,  \quad i \in \{QG\},\\
\Delta D^{min}_i \leq \Delta D_i \leq \Delta D^{max}_i, \,  \quad i \in \{PL\},
\end{array}
\right.
\]
where  $C_i^{RT}$ is the generation cost at bus $i$ in $(\$/MWh)$\footnote{This price can be the same as day--ahead market or can be changed by the generator in a specific time (i.e. 4P.M. -- 6P.M. in PJM market).}, $\Delta G_i$ is the change in the output of generator $i$, and $\Delta D_i$ is the change in the demand of dispatchable load at bus $i$ in $(MWh)$, $\Delta G^{max}_i$ and $\Delta G^{min}_i$ are the upper and lower bands for change in the generation of each qualified generator (QG)\footnote{All PJM generation units that are following PJM dispatch instructions,
are eligible to participate in the real--time market (to set the real--time LMP values), these generation units are called qualified generators.}. Similarly,  $\Delta D^{max}_i$  and $\Delta D^{max}_i$ are the upper and lower bands for change in the consumption of each dispatchable load (DL).
Second constraint shows that any change in the transmitted power in congested lines (CL), should be non--positive value.

Similar to day--ahead market, LMP in bus $i$ (without considering the effect of losses) will be,
\begin{equation}
LMP^{RT}_i = \lambda + \sum\limits_{i=1}^{L} GSF_{k-i}\times \mu_k,\label{eq:LMP_EXPOST}
\end{equation}
where, $L$ is the number of lines, $\lambda$ is the Lagrangian multiplier of the equality constraint, and $\mu_k$ is the Lagrangian multiplier of the $k^{th}$ transmission constraint.

\subsection{Cyber Attack Against Electricity Prices}\label{Cyber:formu}
 Real-time market uses the state estimator results that shows the on-line state of the network. In order to transfer data to the state estimator, control center uses different communication channels such as power line communication channel. Using these channels, increases the risk of cyber attack. In other word, if an attacker can change the measurement values\footnote{Attacker can carry out stealth attacks by corrupting
the power flow measurements through attacking the Remote Terminal Units (RTUs), tampering with the
heterogeneous communication network or breaking into the Supervisory Control and Data Acquisition (SCADA) system through
the control center office Local Area Network (LAN) \cite{ATSA-2010,Liu}.}, the results of state estimation and consequently results of real-time market will be affected. Changing measurements' data without detection by BDD (which can bring financial benefits) is the main goal of the attacker in this paper. In the previous section, we described that the congestion in lines will change the price of electricity in the network. Manipulating prices is a good incentive for the attacker to compromise the measurements. In order to manipulate the congestion level in a specific line, the attacker needs to define the group of measurements that can increase or decrease the congestion, then the attacker can insert false data into the measurements.
 Equation (\ref{active.power.flow}), shows that any change in voltage angle can change the transmitted power through the line. For example, any increase/decrease in  $\triangle \hat\theta=(\hat\theta_i-\hat\theta_j)$ will increase/decrease the transmitted power. In online monitoring of power systems, the transmitted power from bus $i$ to bus $j$ can be estimated with $\hat{P_{ij}}=\frac{\hat\theta_i-\hat\theta_j}{X_{ij}}$, and this equation together with equation (\ref{eq:state_infer}) gives the following:
\begin{align} \label{incdec}
\hat{P_{ij}}&=\frac{\hat\theta_i-\hat\theta_j}{X_{ij}}=\frac{(\mathbf{M}_i-\mathbf{M}_j)^T}{X_{ij}}\mathbf{z}\\
&=\mathbf{Q}^T\mathbf{z}=\mathbf{Q}_+^T\mathbf{z}_++\mathbf{Q}_-^T\mathbf{z}_-\nonumber,
\end{align}
where $\mathbf{Q}^T=\frac{(\mathbf{M}_i-\mathbf{M}_j)^T}{X_{ij}}$. The positive and negative arrays of this vector are shown with $\mathbf{Q}_+^T$ and $\mathbf{Q}_-^T$, respectively. These coefficient vectors divide the measurements into two groups $\mathbf{z}_+$ and $\mathbf{z}_-$, in which adding $z^a>0$ to any array of $\mathbf{z}_+$ and $\mathbf{z}_-$ will increase and decrease the estimated transmitted power flow, respectively.
 In this paper, the measurements in $\mathbf{z}_+$ and $\mathbf{z}_-$ are considered as group $\mathcal{M}$ and $\mathcal{N}$, respectively\footnote{It is assumed that attacker knows $\mathbf{H}$ (and consequently $\mathbf{M}$). Knowing the location of attack, from (\ref{incdec}), attacker can distinguish the measurements in group $\mathcal{M}$ and $\mathcal{N}$. }.
After defining these groups, the attacker tries to insert an undetectable bad data into the measurements. Assume $\mathbf{z}=\mathbf{z}_0$ is the measurement values without corruption (safe mode). From (\ref{EQ:resid}) residue for safe mode will be:
\begin{equation}
\rr_0 = \z - \HH\hat{\mathbf{\theta}}=\z_0 - \HH (\mathbf{M}\mathbf{z}_0).
\end{equation}
In the case of attack, $\mathbf{z}=\mathbf{z}_0+\mathbf{z}^a$ and the residue will be,

\begin{alignat}{1}
\rr &= \z - \HH\hat{\mathbf{\theta}}=\z_0+\z^a - \HH (\mathbf{M}\mathbf{z}_0+\mathbf{M}\mathbf{z}^a)\\\nonumber
&= \z_0 - \HH \mathbf{M}\mathbf{z}_0+\mathbf{z}^a- \HH \mathbf{M}\mathbf{z}^a=\rr_0+\rr^a,\\\nonumber
\end{alignat}

where $\rr^a=(\mathbf{I}-\mathbf{HM})z^a$. From triangular inequality,
\begin{equation}
\parallel\rr\parallel \leq  \parallel \rr_0 \parallel+\parallel \rr^a \parallel,
\end{equation}
this equation shows that if $\parallel\rr^a\parallel=\parallel(\mathbf{I}-\mathbf{HM})z^a\parallel$ is small, with large probability control center can not distinguish between $\parallel \rr \parallel$ and $\parallel \rr_0 \parallel$. So inserted attack will path the bad data detection if, $\|(\mathbf{I}-\mathbf{HM})\mathbf{z}^a\|\leq \mathbf{\xi}$. In this constraint $\xi$ is a design parameter for the attacker. Smaller values of $\xi$
will be more likely to be undetected by the control center \cite{xie_li}. However, the ability to
manipulate the state estimation, will be limited. we assume $\xi$ is predetermined by the attacker. In order to change congestion, attacker will define the inserted false data using the following optimization,
\begin{eqnarray}\label{ATTACK}
\max_{\mathbf{z^a}}. \quad  \quad \sum\limits_{i\in \{\mathcal{M}\}} z^a(i)- \sum\limits_{j\in \{\mathcal{N}\}} z^a(j),
\end{eqnarray}
\[\mbox{s.t.}\ \left\{
\begin{array}{l l}
\|(\mathbf{I}-\mathbf{HM})\mathbf{z}^a\|\leq \mathbf{\xi},\\
z^a(k)=0 \quad \forall \ {k\in \{\mathcal{SM}\}},
\end{array}
\right.
\]
where $z^a(i)$ is the $i^{th}$ element of attack vector $\mathbf{z}^a$. Group $\mathcal{M}$ and $\mathcal{N}$  consist of measurements that increasing and decreasing their value will increase the congestion. Objective of the above optimization is to increase and decrease measurements value in group $\mathcal{M}$ and $\mathcal{N}$, respectively. First constraint is for avoiding detection of the attack by bad data detector in state estimator.  Group $\mathcal{SM}$ shows the safe measurements that can not be compromised (such as those protected by Phasor Measurement Units). With inserting the resulted attack vector $z^a$ to the actual values of measurements ($\mathbf{z}=\mathbf{z}_0+\mathbf{z}^a$), the attacker will change the estimated transmitted power in the attacked line.
 From (\ref{incdec}), this change will be
\begin{equation}\label{eq:PL_change}
 \Delta \hat{P_{ij}}= \frac{(\mathbf{M}_i-\mathbf{M}_j)^T}{X_{ij}}\mathbf{z}^a.
 \end{equation}
 While the attacker tries to increase this change, the defender tries to decrease it by defending the measurements that have high risk of being attacked. Changing the estimated power flow in a specific line will increase the chance of changing prices in both sides of the attacked line\footnote{The attacker doesn't have access to all data such as the submitted prices, generation limits, etc. So with changing the estimated transmitted power desired direction, the attacker increases the chance of creating or releasing congestion in the attacked line.}.
Either increasing or decreasing congestion can bring financial benefits for attacker.
\subsubsection{Decreasing The Congestion}
In day--ahead market the attacker buys at lower price $LMP_i^{DA}$ and sells at higher price $LMP_j^{DA}$ ($LMP_i^{DA} < LMP_j^{DA}$). The difference of two prices should be paid to the transmission company as the congestion prices. In the real--time market, because of decreasing congestion, the congestion price paid by the attacker is less than the supposed congestion price in the day--ahead market so the profit of this trade in \$$/MWh$ will be:
\begin{align} \label{PRO_DEC}
P_{Cng}^{Dec}&=Congestion_{Price}^{DA}-Congestion_{Price}^{RT}\\
&=(LMP_j^{DA}-LMP_i^{DA})-(LMP_j^{RT}-LMP_i^{RT}\nonumber).
\end{align}

\subsubsection{Increasing the congestion} Increasing transmitted power from bus $i$ to bus $j$, can create congestion in line $L_{ij}$. This congestion increases/decreases the price of electricity in the receiving/sending end of the transmission line. So the attacker needs to buy a Financial Transmission Right (FTR) from sending bus $i$ to ending bus $j$. FTR is a financial contract  to hedge congestion charges. The FTR holder has access to a specific transmission line in a defined time and location to transmit a specific value of power. In real--time market with creating congestion, FTR can be sold (with higher price) to any Load Serving Entities (LSE's).

In the next section, we will analyze the behavior of both attacker and defender in the real--time market. Limitation in attack (to) and defend (from) different measurements makes a difficult situation for both parties. Mathematical modeling of this behavior in the next section, is an efficient answer to the question of \emph{where should I attack?} and \emph{where should I defend?} for the attacker and the defender, respectively.

\section{Gaming Between Attacker and Defender}\label{sec:game}
In order to protect line $L$, the defender needs to protect group $\mathcal{M}$ and group $\mathcal{N}$. Because the inserted attack will pass the BDD in state estimation (first constraint in (\ref{ATTACK})), the control center should use some other detection methods. For example, the defender can put some secure measurements into random locations in the network. The main problem in this procedure is that defending all measurements is not possible. On the other hand, it is impossible for the attacker to attack all measurements. Instead it tries to attack measurements that have the most effect on the state estimator without being detected by the control center.  This behavior can be modeled with a zero--sum strategic game between the attacker and the defender\footnote{In the case that there are different non-cooperative attackers, they will have the worst performance. But if the attackers are cooperative, it is the worst case for the defender. In this paper, we consider the worst case by assuming all attackers are together as one party. So we formulate the problem as the two-user zero sum game. If the attackers are non-cooperative, some games such as the Stackelberg game can be employed. These games are interesting topics which needs future investigations.}.
\subsection{Two-Person Zero-Sum Game Between Attacker and Defender}

 Define  $A=(\mathcal{N},(\mathcal{S}_i)_{i\epsilon \mathcal{R}},(\mathcal{U}_i)_{i\epsilon \mathcal{N}})$ as a game, in which the defender and the attacker compete to increase and decrease the change of the estimated transmitted power ($\Delta \hat{P_{ij}}$), respectively. In this game, $\mathcal{R}$ is the set of players (the defender and the attacker), and the game can be defined as:

\begin{itemize}
\item Players set: $\mathcal{R}=\{1,2\}$ (the defender and the attacker).\\

 \item  Attacker's strategy: to choose measurements to attack.\\

 \item Strategy set $\mathcal{S}_i$:  The set of available strategies for player $i$, $\mathcal{S}_1=\{{_{\alpha}}C_{N_a}\},$ $\mathcal{S}_2=\{{_{\alpha}}C_{N_d} \},$ where $N_a$ and $N_d$ are the maximum number of measurements that the attacker can attack and the defender can defend and ${_{\alpha}}C_{N_a}$ is the combination of $N_a$ measurement out of $\alpha$ measurement.

    \item Utility: $U_1=\Delta \hat{P_{ij}}$ and $U_2=-\Delta \hat{P_{ij}}$ for the attacker and the defender, respectively.

 \end{itemize}

\subsection{Noncooperative Finite Games: Two--Person Zero--Sum}
A strategic game is a model of interactive decision-making, in which each decision-maker chooses its plan of action once and for all, and these choices are made simultaneously.
For a given $(m \times n)$ matrix game $\mathbf{A} = \{ a_{ij}: i = 1, \dots ,m;j = 1,\dots,n \}$, let $\{ row \; i^*, column \; j^* \}$ be a pair of strategies adopted by the players. Then, if the pair of inequalities
\begin{equation}
a_{i^*j} \leq a_{i^*j^*} \leq a_{ij^*},\label{Single_saddle}
\end{equation}
is satisfied $\forall i, j$. The two--person zero--sum game is said to have a saddle point in pure strategies. The strategies \{row $i^*$, column $j^*$\} are said to constitute a saddle--point equilibrium. Or simply, they are said to be the saddle--point strategies. The corresponding outcome $a_{i^*j^*}$ of the game is called the saddle--point value.
If a two--person zero--sun game possesses a single saddle point, the value of the game is uniquely given by the value of saddle point. However, the mixed strategies are used to obtain an equilibrium solution in the matrix games that do not possess a saddle point in pure strategies. A mixed strategy for a player is a probability distribution on the space of its pure strategies.
Given an $(m $$\times$$ n)$ matrix game $\mathbf{A} = \{ a_{ij}: i = 1, \dots ,m;j = 1,\dots,n \}$, the frequencies with which different rows and columns of the matrix are chosen by the defender and the attacker will converge to their respective probability distributions that characterize the strategies. In this way, the average value of the outcome of the game is equal to
\begin{equation}
J(\mathbf{y},\mathbf{w}) = \sum\limits_{i=1}^{m} \sum\limits_{j=1}^{n} y_i a_{ij} w_j = \mathbf{y'}\mathbf{A}\mathbf{w},
\end{equation}
where $\mathbf{y}$ and $\mathbf{w}$ are the probability distribution vectors defined by
\begin{equation}
\mathbf{y} = (y_1,\cdots,y_m)' ,\quad \mathbf{w} = (w_1,\cdots,w_n)'.
\end{equation}
The defender wants to minimize $J(\mathbf{y},\mathbf{w})$ by an optimum choice of a probability distribution vector $\mathbf{y} \in Y$, while the attacker wants to maximize the same quantity by choosing an appropriate $\mathbf{w} \in W$. The sets $Y$ and $W$ are
\begin{equation}
Y = \{ \mathbf{y} \in R^m : \mathbf{y} \geq \mathbf{0},\quad \sum\limits_{i=1}^{m} y_i = 1\},
\end{equation}
\begin{equation}
W = \{ \mathbf{w} \in R^n : \mathbf{w} \geq \mathbf{0},\quad \sum\limits_{j=1}^{n} w_j = 1\}.
\end{equation}
Given an $(m $$\times$$ n)$ matrix game $\mathbf{A}$, a vector $\mathbf{y^*}$ is known as a mixed security strategy for the defender if the following inequality holds $\forall\mathbf{y} \in Y$:
\begin{equation}
\overline V_m(\mathbf{A}) \triangleq \max \limits_{\mathbf{w} \in W} \mathbf{{y^{*}}}'\mathbf{A}\mathbf{w} \leq \max \limits_{\mathbf{w} \in W} \mathbf{y'}\mathbf{A}\mathbf{w}, \quad  \mathbf{y} \in Y.
\end{equation}
And the quantity $\overline V_m(\mathbf{A})$ is known as the average security level of the defender. We can also define the average security level of the attacker as $\underline V_m(\mathbf{A})$ if the following inequality holds for all $\mathbf{w} \in W$:
\begin{equation}
\underline V_m(\mathbf{A}) \triangleq \min \limits_{\mathbf{y} \in Y} \mathbf{y'}\mathbf{A}\mathbf{w^*} \geq \min \limits_{\mathbf{y} \in Y} \mathbf{y'}\mathbf{A}\mathbf{w}, \quad \mathbf{w} \in W.
\end{equation}
The two inequalities can also be given as:
\begin{equation}
\overline V_m(\mathbf{A}) = \min \limits_Y \max \limits_W \mathbf{y'}\mathbf{A}\mathbf{w},
\end{equation}
\begin{equation}
\underline V_m(\mathbf{A}) = \max \limits_W \min \limits_Y \mathbf{y'}\mathbf{A}\mathbf{w}.
\end{equation}
However, it always holds true that $\underline V_m(\mathbf{A}) = \overline V_m(\mathbf{A})$ for a two-person zero-sum game in the mixed strategies. In this way, for an $(m $$\times$$ n)$ matrix game $\mathbf{A}$, $\mathbf{A}$ has a saddle point in the mixed strategies, and $V_m(\mathbf{A})$ is uniquely given by
\begin{equation}
V_m(\mathbf{A}) = \overline V_m(\mathbf{A}) = \underline V_m(\mathbf{A}).
\end{equation}
 We can see that if the players are able to use mixed strategies, the matrix games always have a saddle-point solution $V_m(\mathbf{A})$ as the only solution in the zero-sum two-person game.

%

\subsection{Computation of A Two-Person Zero-Sum Game}

One way to get the saddle point in the mixed strategies is to convert the original matrix game into a linear programming (LP) problem. Given  $\mathbf{A} = \{ a_{ij}: i = 1, \dots ,m;j = 1,\dots,n \}$ with all entries positive $(i.e., a_{ij}>0)$, the average value of the game in mixed strategies is given by
\begin{equation}
V_m(\mathbf{A}) = \min \limits_Y \max \limits_W \mathbf{y'}\mathbf{A}\mathbf{w} = \max \limits_W \min \limits_Y \mathbf{y'}\mathbf{A}\mathbf{w}.
\end{equation}
Obviously, $V_m(\mathbf{A})$ must be a positive quantity on $\mathbf{A}$. Furthermore, the expression can also be written as
\begin{equation}
\min \limits_{\mathbf{y} \in Y} v_1(\mathbf{y}),
\end{equation}
where $v_1(\mathbf{y})$ is defined as
\begin{equation}
v_1(\mathbf{y}) = \max \limits_W \mathbf{y'}\mathbf{A}\mathbf{w} \geq \mathbf{y'}\mathbf{A}\mathbf{w}, \quad \forall \mathbf{w} \in W.
\end{equation}
In addition, it can also be written as
\begin{equation}
\mathbf{A'}\mathbf{y} \leq \mathbf{1_n}v_1(\mathbf{y}),\quad \mathbf{1_n} \triangleq {(1, \dots ,1)}' \in R^n.
\end{equation}
Now the mixed security strategy for the defender is to
\begin{eqnarray}
\min  \ v_1(\mathbf{y})
\end{eqnarray}
\[\mbox{s.t.}\
\left\{
\begin{array}{l l}
A' \tilde{\mathbf{y}} \leq \mathbf{1_n},\\
{\tilde{\mathbf{y}}}'\mathbf{1_m} = [v_1(\mathbf{y})]^{-1},\\
\mathbf{y} = \tilde{\mathbf{y}}v_1(\mathbf{y})\\
\tilde{\mathbf{y}} \geq 0,
\end{array}
\right.
\]
where $\tilde{\mathbf{y}}$ is defined as $\mathbf{y}/v_1(\mathbf{y})$. This is further equivalent to the maximization problem
\begin{eqnarray}
\max \limits_{\mathbf{\tilde{y}}} \ \mathbf{{\tilde{y}}'1_m},\label{eq:def}
\end{eqnarray}
\[\mbox{s.t.}\
\left\{
\begin{array}{l l}
\mathbf{A'}\tilde{\mathbf{y}}\leq \mathbf{1_n},\\
\tilde{\mathbf{y}} \geq 0,
\end{array}
\right.
\]
which is a standard LP problem.

Similarly, we can get the standard LP problem for the attacker
\begin{eqnarray}
\min\limits_{\mathbf{\tilde{w}}} \ {\tilde{\mathbf{w}}}'\mathbf{1_n}, \label{eq:att} \quad
\end{eqnarray}
\[\mbox{s.t.}\ \left\{
\begin{array}{l l}
\mathbf{A}\tilde{\mathbf{w}}\geq \mathbf{1_m},\\
\tilde{\mathbf{w}} \geq 0,
\end{array}
\right.
\]
where $\mathbf{\tilde{w}}$ is defined as $\mathbf{w}/v_2(\mathbf{w})$ and
\begin{equation}
v_2 \triangleq \min \limits_Y \mathbf{y'}\mathbf{A}\mathbf{w} \leq \mathbf{y'}\mathbf{A}\mathbf{w}, \quad \forall \mathbf{y} \in Y.
\end{equation}


\begin{table}[ht]
\begin{center}
\caption{Line Reactance and thermal limit for 5--bus test system}
\begin{tabular}{|c|c|c|c|c|c|c|}
  \hline
  Line & $L_{12}$ & $L_{14}$ & $L_{15}$ & $L_{23}$& $L_{34}$ & $L_{45}$ \\
  \hline
  X (\%) & 2.81 & 3.04 & 0.64 & 1.08 & 2.97 & 2.97 \\
  \hline
  $F_{k}^{max}(MW)$ & 999 & 999 & 999 & 999 & 999 & 240 \\
  \hline
\end{tabular}
\label{table:line_param}
\end{center}
\end{table}

\section{Numerical Results}\label{res:list}

\begin{figure}
\centering
\includegraphics[width=90mm]{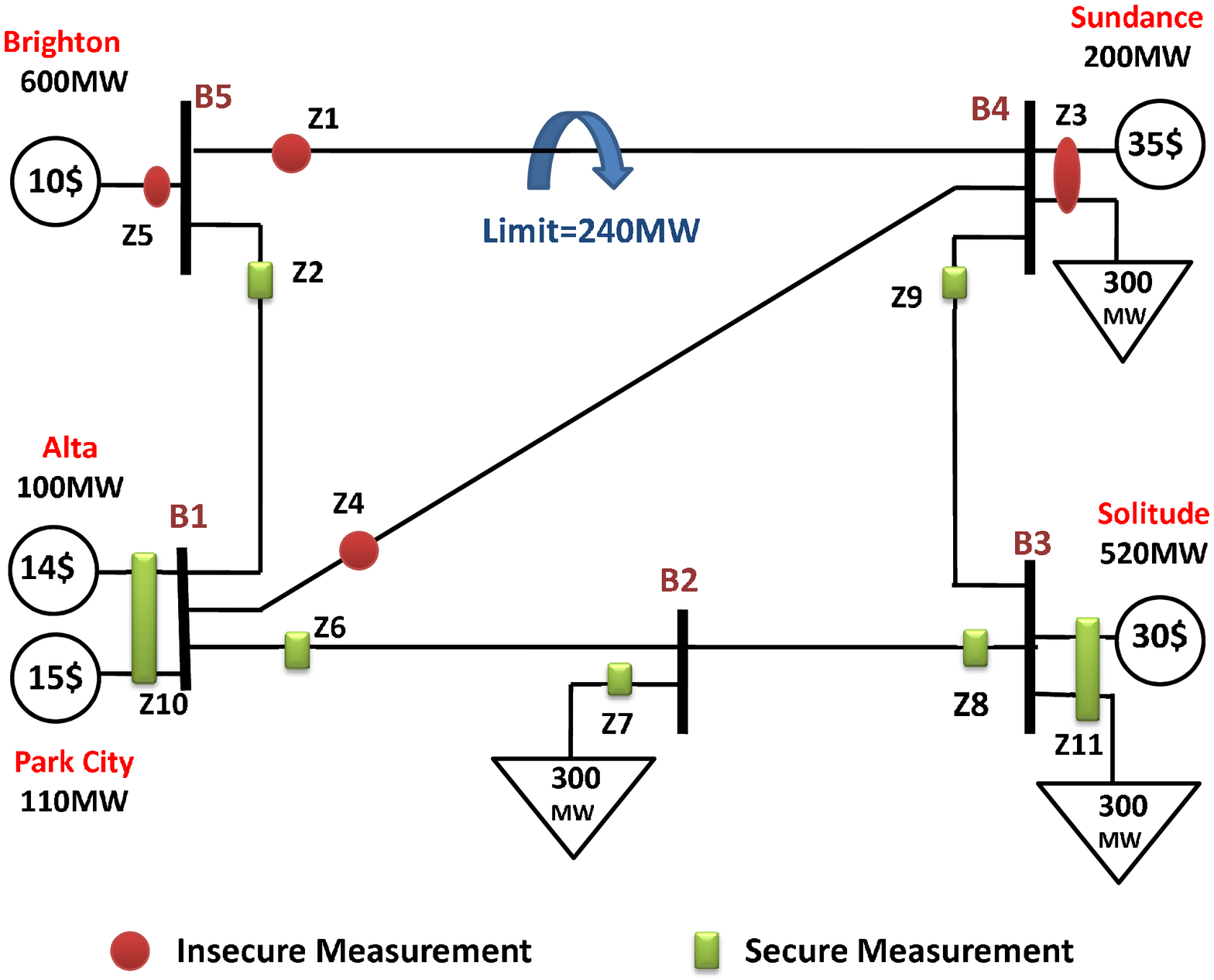}
\caption{Measurement configuration in PJM 5-bus test system}
\label{f: system_model}
\end{figure}

\begin{figure}[th]
\begin{center}
\includegraphics[width=3.8in]{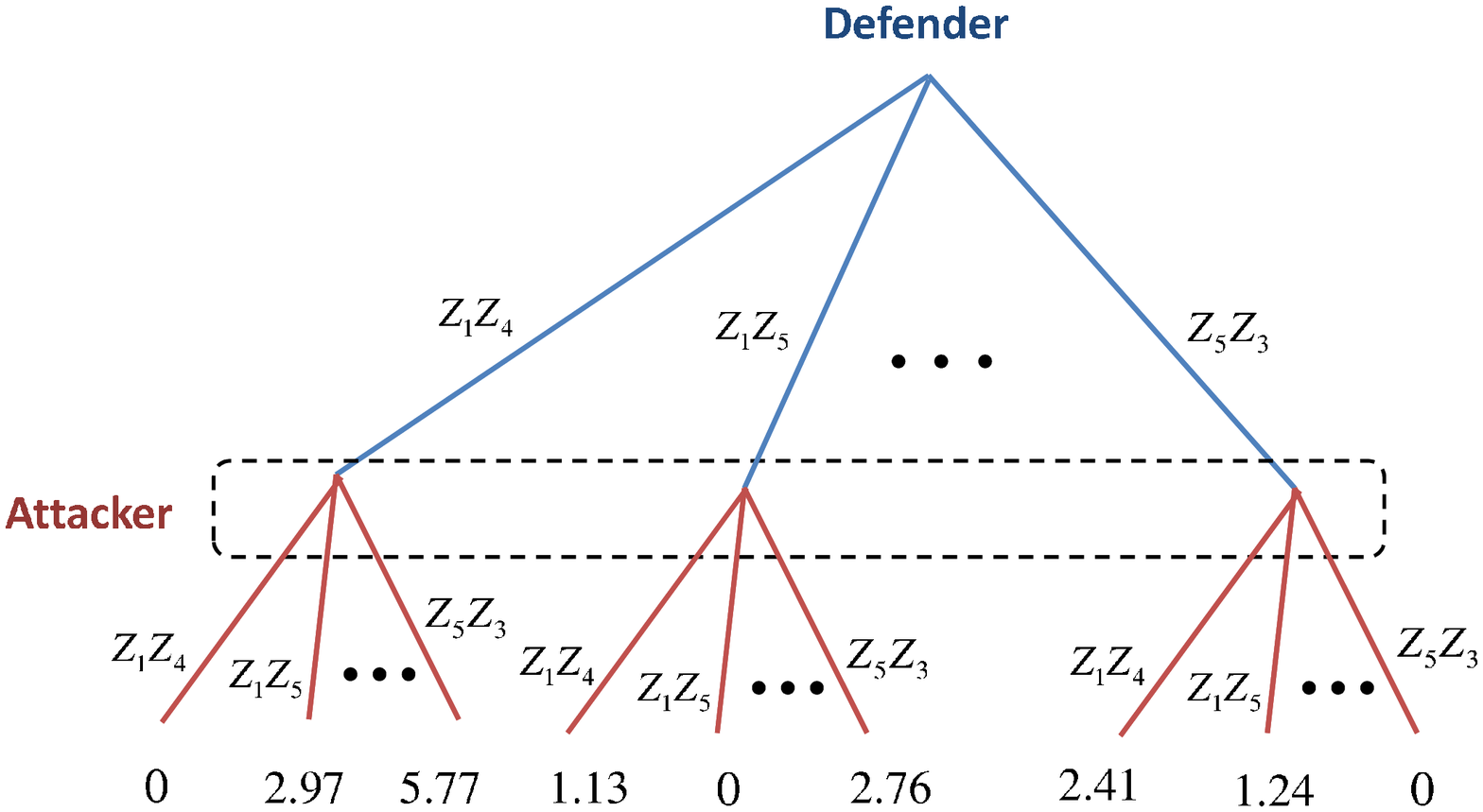}\vspace{-23mm}
\end{center}
\caption{Extensive form of single--act game}
\label{fig:single_act_game}
\end{figure}

In this section, we analyze the effect of attack on the PJM 5-bus test system in \cite{PJM_ref} with a slightly modifications. Transmission lines' parameters are given in Table \ref{table:line_param} and \ref{table:GSF}. Generators' and loads' parameters (including $G_i^{max}$, $C_i$, and $D_i$) together with the location of measurements are shown in
Figure \ref{f: system_model}. Solving (\ref{DCOPFNL_EA}) for the day--ahead market shows that $L_{54}$ (line from $B_5$ to $B_4$) is congested. Here attacker chooses $L_{54}$ to attack. Knowing $H$, from (\ref{incdec}) the attacker obtains $\mathbf{Q}=[0.2\    0.05\    0\    0.19\    0.25\    0.04\   -0.04\   -0.08\   -0.13\    0.18\    0.05] $. Positive and negative arrays of this vector correspond to $z_+$ and $z_-$ vectors, respectively, i.e., $z_+^T=[z_1,z_2,z_4,z_5,z_6,z_{10}]$ and $z_-^T=[z_7,z_8,z_9] $. The greater values of $Q(i)$ correspond to measurements that have more effect on $\hat{P_{ij}}$. Suppose there are 4 insecure measurements $\{z_1,z_4,z_5,z_{10}\}$ and the attacker can compromise 2 of them, also the defender can defend 2 measurements simultaneously. So the attacker should choose 2 measurements among these measurements that have more effect on $\hat{P_{ij}}$ and a sufficiently low probability of detection by the defender. In this example, the attacker can choose from strategy set  $\mathcal{S}_1=\{ z_1z_4, z_1z_5, z_1z_{3}, z_4z_5, z_4z_{3}, z_5z_{3}\}$, and the defender can choose from strategy set $\mathcal{S}_2=\{z_1z_5, z_1z_{3}, z_4z_5, z_4z_{3}, z_5z_{3}\}$. It is assumed that if the attacker for example chooses $\{ z_iz_j\}$ (to attack measurement $i$ and $j$, $i\neq j$) and the defender chooses $\{ z_iz_k\}$ (to defend measurement $i$ and $k$, $i\neq k$), compromising $\{ z_j\}$ will be successful, and the change in $\hat{P_{ij}}$ is only because of compromising $\{ z_j\}$. If $\mathbf{\xi}=[5_{MW},\cdots,5_{MW}]'_{(12\times 1)}$, solving (\ref{ATTACK}) and (\ref{eq:PL_change}) gives $\Delta \hat P_{54}=U_1=-U_2$. As Figure \ref{fig:single_act_game} shows, these payoffs are the results of different attack and defend strategies (which both players take). The attacker and defender in this game are not aware of the sequence of play. Also one player has no idea about the other player's action. These situations are described by a normal form zero--sum game in Table \ref{table:Game}.

\begin{table}[ht]
\begin{center}
\caption{Generation shift factors of lines in 5--bus test system}
\begin{tabular}{|c|c|c|c|c|c|c|}
  \hline
\backslashbox{Line}{Bus}&  $B_1$ &	$B_2$ &	$B_3$ &	 $B_4$&	 $B_5$\\
\hline
$L_{1-2}$&  0.1939&	-0.476&	-0.349&	0&	0.1595\\
$L_{1-4}$&  0.4376&	0.258 &	0.1895&	0&	0.36\\
$L_{1-5}$&  0.3685&	0.2176&	0.1595&	0& -0.5195\\
$L_{2-3}$&  0.1939&	0.5241&	-0.349&	0&	0.1595\\
$L_{3-4}$&  0.1939&	0.5241&	0.6510&	0&	0.1595\\
$L_{5-4}$&  0.3685&	0.2176&	0.1595&	0&	0.4805\\
\hline
\end{tabular}
\label{table:GSF}
\end{center}
\end{table}

\begin{table}[ht]
\caption{zero--sum game between the Attacker and the Defender}
\begin{tabular}{c|c|c|c|c|c|c|c|}
\multicolumn{1}{c} {  } & \multicolumn{1}{c} {   }& \multicolumn{1}{c} { $w_1$ }& \multicolumn{1}{c} { $w_2$ }& \multicolumn{1}{c} { $w_3$ } & \multicolumn{1}{c} { $w_4$ }& \multicolumn{1}{c} { $w_5$ }& \multicolumn{1}{c} { $w_6$ }\\
  \cline{2-8}
   & \backslashbox{Def.}{Att.} & $z_1z_4$ & $z_1z_5$ & $z_1z_{10}$ & $z_4z_5$ & $z_4z_{10}$ & $z_5z_{10}$ \\
  \cline{2-8}
  $y_1$ &  $z_1z_4$    & 0&	3.14&	2.81&	3.14&	 2.81&	 4.84  \\
  \cline{2-8}
  $y_2$ &  $z_1z_5$    &  1.17 &	0	&2.81&	1.17&	 5&	 2.81 \\
  \cline{2-8}
  $y_3$ &  $z_1z_{10}$ &1.17	&3.14&	0&	5  &	1.17&	 3.14 \\
  \cline{2-8}
  $y_4$ &   $z_4z_5$   & 1.28&	1.28&	4.43&	0&	 2.81&	 2.81 \\
  \cline{2-8}
  $y_5$ &   $z_4z_{10}$& 1.28&	5.35&	1.28&	3.14&	 0&	 3.14\\
  \cline{2-8}
  $y_6$ &  $z_5z_{10}$ & 3.21 &	1.28&	1.28&	1.17&	 1.17&	0 \\
  \cline{2-8}
\end{tabular}
 \label{table:Game}
\end{table}

Table \ref{table:Game} shows that $\min(\max \limits_{row})=3.21$, which is not equal to $\max(\min \limits_{column})=0$. So there is no $a_{i^*j^*}$ that satisfies (\ref{Single_saddle}). Therefore, the game doesn't have a single saddle point and the problem shifts to finding the proportion of times that the attacker and the defender, play their own strategies. Solving such a game (which does not have a single saddle point) is a linear programming. From (\ref{eq:def}) defender defines $\mathbf{{\tilde{y}}}$, we have
\begin{eqnarray}\label{Game:Defend}
\max \quad  \mathbf{{\tilde{y}}'1_m},
\end{eqnarray}
\[\mbox{s.t.}\ \left\{
\begin{array}{l l}
1.17\tilde{y}_2+1.17\tilde{y}_3+1.28\tilde{y}_4+1.28\tilde{y}_5+3.2\tilde{y}_6\leq 1,\\
3.14\tilde{y}_1+3.14\tilde{y}_3+1.28\tilde{y}_4+5.35\tilde{y}_5+1.28\tilde{y}_6\leq 1,\\
2.81\tilde{y}_1+2.81\tilde{y}_2+4.43\tilde{y}_4+1.28\tilde{y}_5+1.28\tilde{y}_6\leq 1,\\
3.14\tilde{y}_1+1.17\tilde{y}_2+5\tilde{y}_3+3.14\tilde{y}_5+1.17\tilde{y}_6\leq 1,\\
2.81\tilde{y}_1+5\tilde{y}_2+1.17\tilde{y}_3+2.81\tilde{y}_4+1.17\tilde{y}_6\leq 1,\\
4.84\tilde{y}_1+2.81\tilde{y}_2+3.14\tilde{y}_3+2.81\tilde{y}_4+3.14\tilde{y}_5\leq 1,\\

\tilde{y}_1,\tilde{y}_2,\tilde{y}_3,\tilde{y}_4,\tilde{y}_5,\tilde{y}_6 \geq 0,\\
\end{array}
\right.
\]
which gives $\mathbf{\tilde{y}}=[0\ 0.049\ 0.134\ 0.136\ 0.018\ 0.183 ]$. Therefore,
$\mathbf{y} = \tilde{\mathbf{y}}v_1(\mathbf{y})= \tilde{\mathbf{y}}({\tilde{\mathbf{y}}}'\mathbf{1_m})^{-1}=[0\ 0.094\ 0.26\ 0.262\ 0.0347\ 0.35]$. Similarly, solving (\ref{eq:att}) for the attacker gives $\mathbf{\tilde{w}}=[0.29\ 0\ 0.02\ 0.019\ 0.019\ 0.174]$, and therefore, $\mathbf{w} = \tilde{\mathbf{w}}v_1(\mathbf{w})= \tilde{\mathbf{w}}({\tilde{\mathbf{w}}}'\mathbf{1_m})^{-1}=[0.556\    0\    0.038\    0.036\ 0.037\    0.333]$.

\begin{figure*}
\centerline{\subfigure[Probability of attack ]{\includegraphics[scale=0.35]{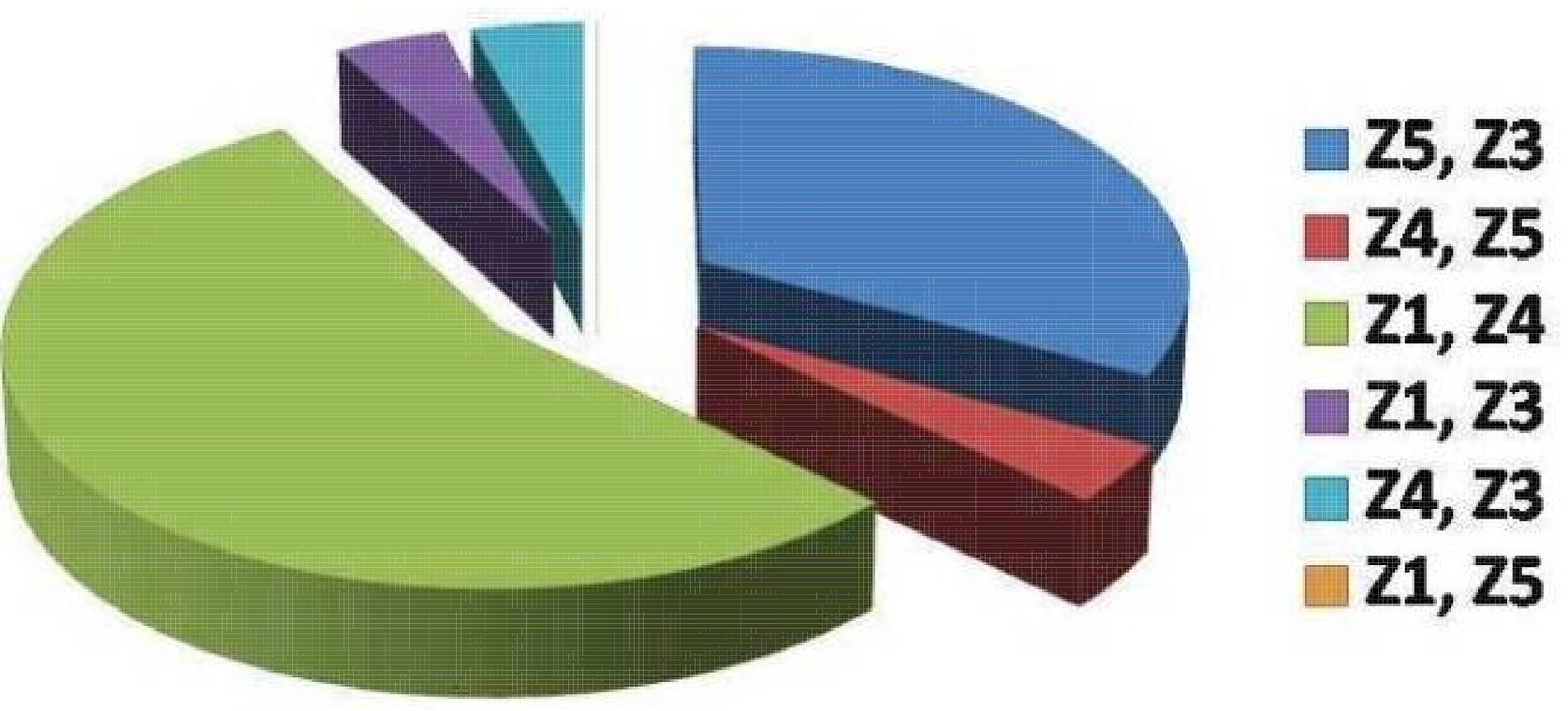}
\label{fig:pattern1}}
\hspace{1cm}
\subfigure[Probability of defend ]{\includegraphics[scale=0.34]{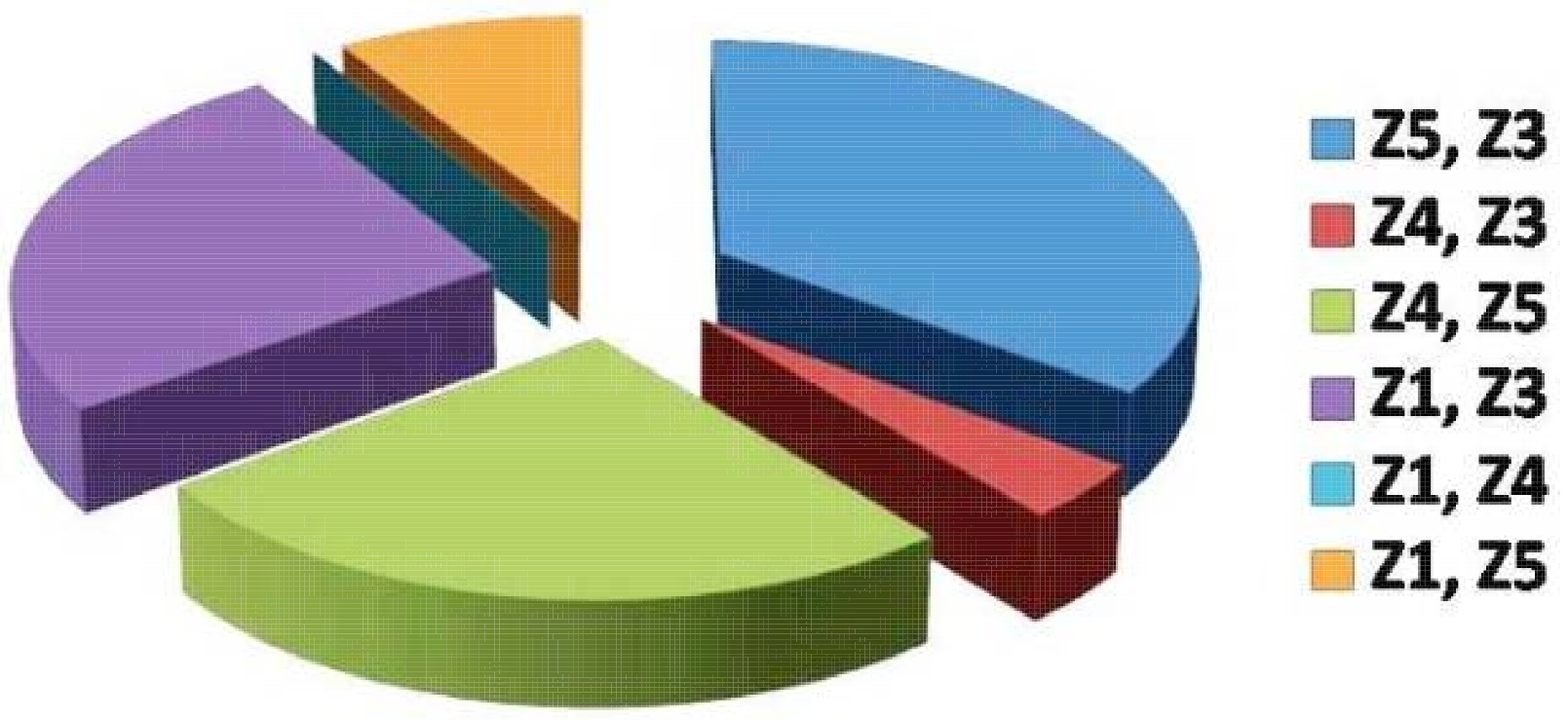}
\label{fig:pattern2}}}
\caption{Proportion of times that attacker and defender, attack and defend to measurements respectively.}
\label{fig:pattern}
\end{figure*}

Figure \ref{fig:pattern} shows the proportion of times that the defender and the attacker should defend and attack different measurements, respectively.
%
%
As discussed in Section \ref{sec:attinmarket}, changing the estimated transmitted power in line $L_{54}$ can change the prices in either bus $5$ or bus $4$. In real--time market the control center estimates transmitted power and then knowing dispatch schedule (which is defined in day--ahead market) load level in different buses is estimated. This estimated load together with the current state of the network is applied to a DCOPF, and this program defines the real--time prices. If the operating condition (such as the load level) has not changed and there is no error in the measurements, the real--time prices should be the same as the day--ahead prices. Here without loss of generality, we assume that the actual load level doesn't change and any change in the estimated load level is because of bad data injection to the state estimator.

The following example shows how attacker is able to change the prices in real--time market. Suppose attacker compromise $z_1z_4$ and the defender defends $z_5z_{10}$ so, attack against $z_1z_4$ is successful. In this case solving (\ref{ATTACK}) gives $\mathbf{z^a}=[8.21\ 0\ 0\ 8.09\ 0\ 0\ 0\ 0\ 0\ 0\ 0\ 0]_{(MW)}$. So from (\ref{eq:state_infer}), estimated states for all buses will be $\mathbf{\hat \theta}=[50\ 56\ 65\ 01\ 71.6]\times 10^{-3}_{(rad)}.$ Using (\ref{incdec}), estimated transmitted power can be obtained\footnote{This value is considered as the real--time transmitted power in $L_{54}$.} $\hat{P_{54}}= 236.59_{(MW)}$. This power is less than thermal limit of transmission line that shows, congestion in this line is released.  In this case solving (\ref{DCOPFNL_EP}) and (\ref{eq:LMP_EXPOST}) gives the real time prices (here it is assumed that $\Delta G_i^{max}=-\Delta G_i^{min}=0.1_{MW}$ and $\Delta D_i^{max}=-\Delta D_i^{min}=0_{MW}$).

\begin{figure}
\centering
\includegraphics[width=3in]{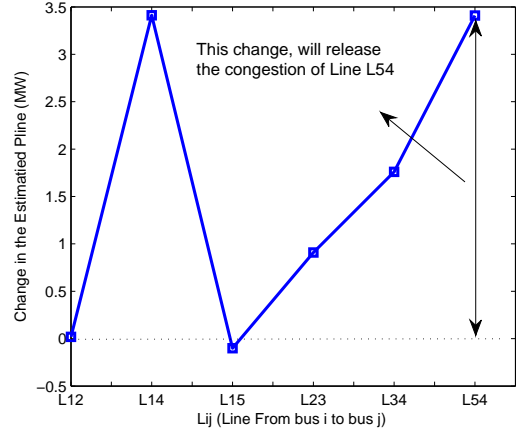}
\caption{Change in the estimated transmitted power of lines because of attack to $Z_1$ and $Z_4$}
\label{fig:del_PLIN}
\end{figure}

\begin{figure}[th]
\begin{center}
\includegraphics[width=3in]{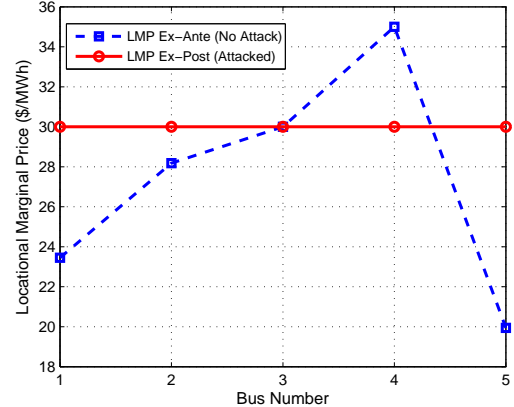}
\end{center}
\caption{Locational Marginal Prices for PJM 5-Bus test system for both with attack and without attack}
\label{fig:LMP}
\end{figure}

Figure \ref{fig:LMP} shows the prices for attacked and without--attack cases. Change of estimated transmitted power in transmission line is shown in Figure \ref{fig:del_PLIN}. Now, assume that in day--ahead market, the attacker buys $100_{MW}$ power in bus $5$ and sells it in bus 4. From (\ref{PRO_DEC}), the profit of this contract will be:
\begin{equation} \label{res:PRO_DEC}
Profit=[(35-20)-(30-30)]\times 100=1500_{(\$/h)}.\\
\end{equation}
%

\section{Conclusion}\label{sec:conclusion}
In this paper, first we analyzed the effect of compromising each measurement on the state estimator results. Compromising these measurements can change the congestion and consequently the price of electricity, and thus, the attacker has an intensive to change the congestion in the desired direction. Since a typical power system has a huge number of measurements, attacking or defending all of those becomes impossible for attacker and defender, respectively. To this end, this behavior is modeled and analyzed in the framework of game theory. The simulation results on PJM 5--Bus test system indicate that, in the specified load level, how attacker can change the prices in the desired direction (decreasing in this example).

\section*{Acknowledgement}

This work is partially supported by US NSF CNS-0953377, ECCS-1028782, CNS-1117560, Qatar National Research Fund, National Nature Science Foundation of China under grant number 60972009 and 61061130561, as well as the National Science and Technology Major Project of China under grant number 2011ZX03005-002.

\bibliographystyle{abbrv}

\begin{thebibliography}{10}

\bibitem{TFGA-2008} T. F. Garrity, ``Getting Smart," {\em IEEE Power and Energy Magazine}, vol. 6, no. 2, pp. 38--45, March--April 2008.

\bibitem{Monti} A. Monticelli, ``Electric Power System State Estimation," {\em Proceedings of IEEE}, vol. 88, no. 2, pp. 262--282, Feb. 2000.


\bibitem{KEN1} Y. Yuan, Z. Li, and K. Ren, ``Modeling Load Redistribution Attacks in Power System," {\em IEEE Transactions on Smart Grid,}, vol. 2, no. 2, pp. 382--390, June 2011.

\bibitem{KEN2} Y. Yuan, Z. Li, and K. Ren, ``Quantitative Analysis of Load Redistribution Attack in Electric Grid," {\em IEEE Transactions on Parallel and Distributed Systems}, vol. 23, no. 9, pp. 1731--1738, Sept. 2012.

\bibitem{Meserve} J. Meserve, ``Staged cyber attack reveals vulnerability in power grid", {\em Available: http://www.cnn.com/2007/US/09/26/power.at.risk/index.html}, CNN, Sep. 2007.

\bibitem{WCNCesmali} M. Esmalifalak, Z. Han, and L. Song ``Effect Of Stealthy Bad Data Injection On Network Congestion In Market Based Power System", {\em IEEE Wireless Communications and Networking Conference (WCNC 2012)}, Paris, France, Apr. 2012.

\bibitem{xie_li} L. Xie, Y. Mo, and B. Sinopoli, ``Integrity Data Attacks in Power Market Operations," {\em IEEE Transactions on Smart Grid}, vol. 2, no. 99, pp. 659--666, Dec. 2011.


\bibitem{Kumar_Srivastava} G. Chen, Z. Y. Dong, D. J. Hill, and Y. S. Xue, ``A Zonal Congestion Management Approach Using Real and Reactive Power Rescheduling," {\em IEEE Transactions on Power Systems}, vol. 19, no. 1, pp. 554--562, Feb. 2004.


\bibitem{MALAesmali} M.E Falak, M.O Buygi, and A. Karimpour, ``Market oriented reactive power expansion planning using locational marginal price", {\em IEEE 2nd International Power and Energy Conference (PECon 2008)}, Johor Baharu, Malaysia, Dec. 2008.

\bibitem{XFSM-} X. Fang. S. Misra, G. Xue, and D. Yang, ``Smart Grid -- The New and Improved Power Grid: A Survey," {\em IEEE Communications Surveys \& Tutorials}, no. 99, pp. 1--37.

\bibitem{E-Brown} H. E. Brown and S. Suryanarayanan, ``A Survey Seeking a Defenition of a Smart Distribution System," {\em North American Power Symposium 2009},  pp. 1--7, 2009.

\bibitem{Rohjansand} S. Rohjansand, M. Uslar, R. Bleiker, J. Gonz$\acute{a}$lez, M. Specht, T. Suding, and T. Weidelt., ``Survey of Smart Grid Standardization Studies and Recommendations," {\em Smart Grid Communications (SmartGridComm), 2010 First IEEE International Conference on}, Oldenburg, Germany, Oct. 2010.

\bibitem{NIST_Roadmap} Office of the National Coordinator for Smart Grid Interoperability, ``NIST Framework and Roadmap for  Smart Grid Interoperability Standards, Release 1.0," {\em Available: http://www.nist.gov/public\_affairs/releases/upload/smartgrid-interoperability\_final.pdf}, Jan. 2010.

\bibitem{ATSA-2010} A. Teixeira, S. Amin, H. Sandberg, K.H. Johansson and S.S. Sastry, ``Cyber security analysis of state estimators in electric power systems," {\em 2010 49th IEEE Conference on Decision and Control (CDC)}, Dec. 2010.

\bibitem{Liu}  Y. Liu, M. K. Reiter, and P. Ning,
``False Data Injection Attacks Against State Estimation in Electric Power Grids," {\em the 16th ACM conference on Computer and communications security.},  Nov. 2009.

\bibitem{Esmali_ICA} M. Esmalifalak, H. Nguyen, R. Zheng, and Z. Han, ``Stealth False Data Injection using Independent Component Analysis in Smart Grid", {\em IEEE Second Conference on Smart Grid Communications}, Brussels, Belgium, Oct. 2011.


\bibitem{AMRE-2010-1} A. R. Metke and R. L. Ekl, ``Smart Grid Security Technology," {\em Innovative Smart Grid Technologies (ISGT), 2010}, Schaumburg, IL, USA, Jan. 2010.

\bibitem{Husheng_Li} H. Li, L. Lai, and R.C. Qiu, ``Communication Capacity Requirement for Reliable and Secure State Estimation in Smart Grid", {\em The first IEEE Conference on Smart Grid Communications}, Maryland, USA, Oct. 2010.

\bibitem{GCZD-2011} G. Chen, Z. Y. Dong, D. J. Hill, and Y. S. Xue, ``Exploring Reliable Strategies for Defending Power Systems Against Targeted Attacks," {\em IEEE Transactions on Power Systems}, vol. 26, no. 3, pp. 1000--1009, Aug. 2011.

\bibitem{NIST_DRAFT} The Smart Grid Interoperability Panel Cyber Security Working Group, ``Introduction to NISTIR 7628 Guidelines for Smart Grid Cyber Security," {\em Available: http://www.nist.gov/smartgrid/upload/nistir-7628\_total.pdf,} Sep. 2010.

\bibitem{Swearingen} M. Swearingen, ``Real Time Evaluation and Operation of the Smart Grid Using Game Theory," {\em 2011 IEEE
Rural Electric Power Conference (REPC)}, Hooker, OK, USA, April 2011.

\bibitem{ZhuLambotharan} Z. Zhu, J. Tang, S. Lambotharan, W.H. Chin, and Z. Fan, ``An Integer Linear Programming and Game
Theory based Optimization for Demand-side Management in Smart Grid," {\em 2011 IEEE GLOBECOM Workshops (GC
Wkshps)}, Loughborough, UK, Dec. 2011.

\bibitem{Fadlullah} Z.M. Fadlullah, Y. Nozaki, A. Takeuchi, and N. Kato, ``A survey of game theoretic approaches in
smart grid," {\em 2011 International Conference on Wireless Communications and Signal Processing (WCSP)},
Sendai, Japan, Nov. 2011.


\bibitem{ZhangLu} X. Zhang, J. Lu, H. Sun, and X. Ma, ``Orderly Consumption and Intelligent Demand-side Response
Management System under Smart Grid," {\em 2010 Asia-Pacific Power and Energy Engineering Conference
(APPEEC)}, Beijing, China, March2010.

\bibitem{WangHuang} P. Wang, J. Y.Huang, Y. Ding, P. Loh, and L. Goel, ``Demand Side Load Management of Smart Grids
using intelligent trading/Metering/ Billing System," {\em 2010 IEEE Power and Energy Society General
Meeting, Singapore}, Singapore, July 2010.

\bibitem{BuYuLiu} S. Bu, F. R.Yu, and P. X.Liu, ``A Game-Theoretical Decision-Making Scheme for Electricity Retailers
in the Smart Grid with Demand-Side Management," {\em 2011 IEEE International Conference on Smart Grid
Communications (SmartGridComm)}, Ottawa, ON, Canada, Oct. 2011.

\bibitem{MohsenianWong} A. Mohsenian-Rad, V.W.S. Wong, J. Jatskevich, R. Schober, and A. Leon-Garcia, ``Autonomous Demand--Side Management Based on Game-Theoretic Energy Consumption Scheduling for the Future Smart Grid," {\em IEEE Transactions on Smart Grid}, vol.1, no.3, pp.320-331, Dec. 2010.



\bibitem{woll} A. J. Wood and B. F. Wollenberg, {\em Power Generation, Operation, and Control}, Wiley New York et al., 1996.

\bibitem{AAAG-2004} A. Abur and A. G. Exposito, {\em Power System State Estimation: Theory and Implementation}, Marcel Dekker, Inc., 2004.







\bibitem{PJM_ref} F. Li, and R. Bo ``Small Test Systems for Power System Economic Studies," {\em Power and Energy Society General Meeting},  Minneapolis, Minnesota USA,  Jul. 2010.

%


\bibitem{FLRB-2007} F. Li and R. Bo, ``DCOPF-Based LMP Simulation: Algorithm, Comparison with ACOPF, and Sensitivity," {\em IEEE Trans. Power Syst.}, vol. 22, no. 4, pp. 1475--1485, Nov. 2007.

\bibitem{FLJP-2004} F. Li, J. Pan, and H. Chao, ``Marginal Loss Calculation in Competitive Electrical Energy Markets," in {\em Proc. 2004 IEEE Int. Conf. Electric Utility Deregulation, Restructuring and Power Technologies 2004 (DRPT 2004)}, Apr. 2004, vol. 1, pp. 205--209.

\bibitem{ELTZ-2004} E. Litvinov, T. Zheng, G. Rosenwald, and P. Shamsollahi, ``Marginal Loss Modeling in LMP Calculation," {\em IEEE Trans. Power Syst.}, vol. 19, no. 2, pp. 880--888, May 2004.

\bibitem{LOTT} A. L. Ott, ``Experience with PJM Market Operation, System Design, and
Implementation," {\em IEEE Trans. Power Syst.}, vol. 18, no. 2, pp. 528–534,
May 2003.
\bibitem{TZHLI} T. Zheng and E. Litvinov, ``Ex post Pricing in the Co-Optimized Energy
and Reserve Market," {\em IEEE Trans. Power Syst.}, vol. 21, no. 4, pp. 1528–1538, Nov. 2006.

\end{thebibliography}


\begin{biography}
[{\includegraphics[width=1in,height
=1.2in,clip,keepaspectratio]{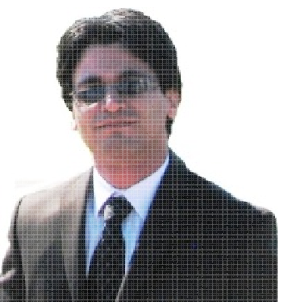}}]{Mohammad Esmalifalak (S'12)}received his M.S. degree in power system engineering from Shahrood University of Technology, Shahrood, Iran in 2007. He joined Ph.D. program in the University of Houston (UH) in 2010. From 2010 to 2012 he was research assistant in the ECE department of UH. He is the author of the paper that won the best paper award in IEEE Wireless Communications and Networking Conference (WCNC 2012), Paris, France. His main research interests include the application of data mining, machine learning and signal processing in the operation and expansion of the smart grids.
\end{biography}
\vspace{-15mm}

\begin{biography}
[{\includegraphics[width=1in,height
=1.2in,clip,keepaspectratio]{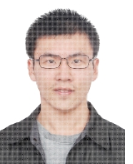}}]{Ge Shi} is a candidate for bachelor's degree of Electronics Engineering from Peking University, Beijing, China. He is now working on a research about intelligent information processing in machine to machine communications (M2M) based on ZigBee protocol under the direction of Prof. Lingyang Song in State Key Laboratory of Advanced Optical Communication Systems \& Networks, Peking University. His research interests mainly include smart grids, game theory, and internet of things.
\end{biography}
\vspace{-15mm}

\begin{biography}
[{\includegraphics[width=.9in,height
=1.2in,clip,keepaspectratio]{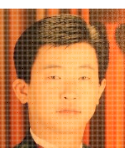}}]{Zhu Han (S'01-M'04-SM'09)} received the B.S. degree in electronic engineering from Tsinghua University, in 1997, and the M.S. and Ph.D. degrees in electrical engineering from the University of Maryland, College Park, in 1999 and 2003, respectively.
92077918

From 2000 to 2002, he was an R\&D Engineer of JDSU, Germantown, Maryland. From 2003 to 2006, he was a Research Associate at the University of Maryland. From 2006 to 2008, he was an assistant professor in Boise State University, Idaho. Currently, he is an Assistant Professor in Electrical and Computer Engineering Department at the University of Houston, Texas. His research interests include wireless resource allocation and management, wireless communications and networking, game theory, wireless multimedia, security, and smart grid communication.

Dr. Han is an Associate Editor of IEEE Transactions on Wireless Communications since 2010. Dr. Han is the winner of IEEE Fred W. Ellersick Prize 2011. Dr. Han is an NSF CAREER award recipient 2010. Dr. Han is the coauthor for the papers that won the best paper awards in IEEE International Conference on Communications 2009, 7th International Symposium on Modeling and Optimization in Mobile, Ad Hoc, and Wireless Networks (WiOpt09), and IEEE Wireless Communication and Networking Conference, 2012.

\end{biography}

\vspace{-20mm}

\begin{IEEEbiography}
[{\includegraphics[width=1in,height
=1.2in,clip,keepaspectratio]{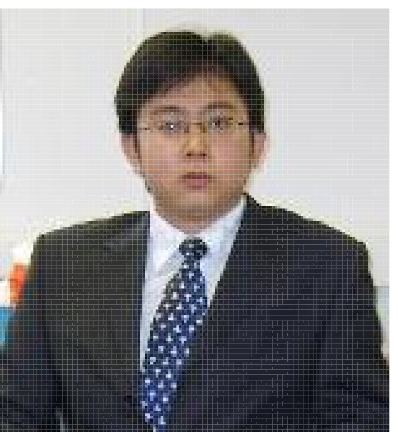}}]{Lingyang Song (S'03-M'06-SM'12)}  received his PhD from the University of York, UK, in 2007, where he received the K. M. Stott Prize for excellent research. He worked as a postdoctoral research fellow at the University of Oslo, Norway, and Harvard University, until rejoining Philips Research UK in March 2008. In May 2009, he joined the School of Electronics Engineering and Computer Science, Peking University, China, as a full professor. His main research interests include MIMO, OFDM, cooperative communications, cognitive radio, physical layer security, game theory, and wireless ad hoc/sensor networks.

He is co-inventor of a number of patents (standard contributions), and author or co-author of over 100 journal and conference papers. He received the best paper award in IEEE International Conference on Wireless Communications, Networking and Mobile Computing (WiCOM 2007), the best paper award in the First IEEE International Conference on Communications in China (ICCC 2012), the best student paper award in the7th International Conference on Communications and Networking in China (ChinaCom2012), and the best paper award in IEEE Wireless Communication and Networking Conference (WCNC2012).

He is currently on the Editorial Board of IET Communications, Journal of Network and Computer Applications, and International Journal of Smart Homes, and a guest editor of Elsevier Computer Communications and EURASIP Journal on Wireless Communications and Networking. He serves as a member of Technical Program Committee and Co-chair for several international conferences and workshops.
\end{IEEEbiography}

\vspace{-8mm}

\end{document}